\documentclass{emulateapj}

\newcommand{\ra}[4]{${#1}^{\rm h}{#2}^{\rm m}{#3}\fs{#4}$}
\newcommand{\dec}[4]{${#1}\arcdeg{#2}\arcmin{#3}\farcs{#4}$}

\slugcomment{Submitted to ApJ.}

\shorttitle{Two Saturn-mass planets}
\shortauthors{Mr\'oz et al.}

\begin{document}

\title{OGLE-2013-BLG-0132Lb and OGLE-2013-BLG-1721Lb: Two Saturn-mass Planets Discovered around M-dwarfs}

\author{Przemek Mr\'oz\altaffilmark{1,18}, A. Udalski\altaffilmark{1}, I.~A. Bond\altaffilmark{2}, J. Skowron\altaffilmark{1}, T. Sumi\altaffilmark{3}, C. Han\altaffilmark{4},}
\affil{}
\author{M.~K. Szyma\'nski\altaffilmark{1}, I. Soszy\'nski\altaffilmark{1}, R. Poleski\altaffilmark{5,1}, P. Pietrukowicz\altaffilmark{1}, \\S. Koz\l{}owski\altaffilmark{1}, \L{}. Wyrzykowski\altaffilmark{1}, K. Ulaczyk\altaffilmark{1,6} \\(The OGLE Collaboration),}
\affil{}
\and
\author{F. Abe\altaffilmark{7}, Y. Asakura\altaffilmark{7}, R.~K. Barry\altaffilmark{8}, D.~P. Bennett\altaffilmark{8}, A. Bhattacharya\altaffilmark{8,9}, M. Donachie\altaffilmark{10}, P. Evans\altaffilmark{10}, A. Fukui\altaffilmark{11}, Y. Hirao\altaffilmark{3}, Y. Itow\altaffilmark{7}, N. Koshimoto\altaffilmark{3}, M.~C.~A. Li\altaffilmark{10}, C.~H. Ling\altaffilmark{2}, K. Masuda\altaffilmark{7}, Y. Matsubara\altaffilmark{7}, Y. Muraki\altaffilmark{7}, M. Nagakane\altaffilmark{3}, K. Ohnishi\altaffilmark{12}, C. Ranc\altaffilmark{8}, N.~J. Rattenbury\altaffilmark{10}, To. Saito\altaffilmark{13}, A. Sharan\altaffilmark{10}, D.~J. Sullivan\altaffilmark{14}, D. Suzuki\altaffilmark{15}, P.~J. Tristram\altaffilmark{16}, T. Yamada\altaffilmark{3}, T. Yamada\altaffilmark{17}, A. Yonehara\altaffilmark{17}\\(The MOA Collaboration)}
\affil{}

\altaffiltext{1}{Warsaw University Observatory, Al. Ujazdowskie 4, 00-478 Warszawa, Poland}
\altaffiltext{2}{Institute for Natural and Mathematical Sciences, Massey University, Private Bag 102904 North Shore Mail Centre, Auckland 0745, New Zealand}
\altaffiltext{3}{Department of Earth and Space Science, Graduate School of Science, Osaka University, 1-1 Machikaneyama, Toyonake, Osaka 560-0043, Japan}
\altaffiltext{4}{Department of Physics, Chungbuk National University, Cheongju 361-763, Korea}
\altaffiltext{5}{Department of Astronomy, Ohio State University, 140 W. 18th Ave., Columbus, OH 43210, USA}
\altaffiltext{6}{Department of Physics, University of Warwick, Coventry CV4 7AL, UK}
\altaffiltext{7}{Institute of Space-Earth Environmental Research, Nagoya University, Furo-cho, Chikusa, Nagoya, Aichi 464-8601, Japan}
\altaffiltext{8}{Laboratory for Exoplanets and Stellar Astrophysics, NASA/Goddard Space Flight Center, Greenbelt, MD 20771, USA}
\altaffiltext{9}{Department of Physics, University of Notre Dame, Notre Dame, IN 46556, USA}
\altaffiltext{10}{Department of Physics, University of Auckland, Private Bag 92019, Auckland, New Zealand}
\altaffiltext{11}{Okayama Astrophysical National Astronomical Observatory, 3037-5 Honjo, Kamogata, Asakuchi, Okayama 719-0232, Japan}
\altaffiltext{12}{Nagano National College of Technology, Nagano 381-8550, Japan}
\altaffiltext{13}{Tokyo Metropolitan College of Industrial Technology, Tokyo 116-8523, Japan}
\altaffiltext{14}{School of Chemical and Physical Sciences, Victoria University, Wellington, New Zealand}
\altaffiltext{15}{Institute of Space and Astronautical Science, Japan Aerospace Exploration Agency, 3-1-1 Yoshinodai, Chuo, Sagamihara, Kanagawa 252-5210, Japan}
\altaffiltext{16}{University of Canterbury Mt John Observatory, PO Box 56, Lake Tekapo 7945, New Zealand}
\altaffiltext{17}{Department of Physics, Faculty of Science, Kyoto Sangyo University, 603-8555 Kyoto, Japan}
\altaffiltext{18}{Corresponding author: pmroz@astrouw.edu.pl}

\begin{abstract}

We present the discovery of two planetary systems consisting of a Saturn-mass planet orbiting an M-dwarf, which were detected in faint microlensing events OGLE-2013-BLG-0132 and OGLE-2013-BLG-1721. The planetary anomalies were covered with high cadence by OGLE and MOA photometric surveys. The light curve modeling indicates that the planet-to-host mass ratios are $(5.15 \pm 0.28)\times 10^{-4}$ and $(13.18 \pm 0.72)\times 10^{-4}$, respectively. Both events were too short and too faint to measure a reliable parallax signal and hence the lens mass. We therefore used a Bayesian analysis to estimate the masses of both planets: $0.29^{+0.16}_{-0.13}\ M_{\rm Jup}$ (OGLE-2013-BLG-0132Lb) and $0.64^{+0.35}_{-0.31}\ M_{\rm Jup}$ (OGLE-2013-BLG-1721Lb). Thanks to a high relative proper motion, OGLE-2013-BLG-0132 is a promising candidate for the high-resolution imaging follow-up. Both planets belong to an increasing sample of sub-Jupiter-mass planets orbiting M-dwarfs beyond the snow line.

\end{abstract}


\section{Introduction}

Gravitational microlensing provides a unique tool for studying the planet formation around late-type stars. The microlensing signal does not depend on the host brightness and the sensitivity of the method happens to peak near or beyond the snow line of majority of the planetary systems. This is a location in the proto-planetary disk where the water ice may condense and where gas giant planets are believed to be formed \citep{mizuno1980,pollack1996}. According to predictions of the core-accretion theory \citep{laughlin2004,ida2005,kennedy2008}, the frequency of gas giant planets should depend on the host mass, with more massive stars more likely to host at least one gas giant. It is also believed that Jovian-mass planets should be relatively rare around M-dwarf stars, while Neptune- and Earth-like planets should be more frequent.

It was recently suggested that sub-Jupiter-mass planets ($0.2<m_{\rm p}/M_{\rm Jup}<1$) around M-dwarfs are more common than Jupiter-mass planets \citep{fukui2015,hirao2016}, which is consistent with predictions of the core-accretion theory. The mass of five such planets is constrained by either high-resolution imaging or parallax: OGLE-2006-BLG-109Lb,c ($0.73\pm0.06\ M_{\rm Jup}$ and $0.27\pm0.03\ M_{\rm Jup}$; \citealt{gaudi2008,bennett2010}), OGLE-2011-BLG-0251Lb ($0.53\pm0.21\ M_{\rm Jup}$; \citealt{kains2013}), OGLE-2011-BLG-0265Lb ($0.9\pm0.3\ M_{\rm Jup}$; \citealt{skowron2015}), and OGLE-2012-BLG-0563Lb ($0.39^{+0.14}_{-0.23}\ M_{\rm Jup}$; \citealt{fukui2015}). For two additional planets, the mass was estimated from the Bayesian analysis: OGLE-2012-BLG-0724Lb ($0.47^{+0.54}_{-0.26}\ M_{\rm Jup}$; \citealt{hirao2016}) and MOA-2010-BLG-353Lb ($0.27^{+0.48}_{-0.16}\ M_{\rm Jup}$; \citealt{rattenbury2015}). However, the number of known planets in this mass regime is still too small to provide strong constraints on theory.

Here, we present the analysis of two faint microlensing events OGLE-2013-BLG-0132 and OGLE-2013-BLG-1721, which were caused by M-dwarfs hosting a Saturn-mass planet. Observations of these events are described in Section \ref{sec:data}. In Section \ref{sec:model}, we discuss the light curve modeling. In Section \ref{sec:source} we characterize the source stars that were magnified during these events, and in Section \ref{sec:bayes}, we infer physical parameters of lenses using the Bayesian analysis. Finally, in Section \ref{sec:conclusions}, we discuss prospects for future follow-up observations.

\section{Data}
\label{sec:data}

Due to their faintness, both events were observed only by the microlensing surveys. The Optical Gravitational Lensing Experiment (OGLE) is monitoring the Galactic bulge using the 1.3~m Warsaw Telescope located at Las Campanas Observatory, Chile. The telescope is equipped with a mosaic 1.4~deg$^2$ CCD camera \citep{udalski2015}. The Microlensing Observations in Astrophysics (MOA; \citealt{bond2001,sumi2013}) group uses the 1.8~m MOA-II telescope at the Mount John University Observatory, New Zealand. The MOA-cam3 CCD camera has a 2.2~deg$^2$ field of view \citep{sako2008}. OGLE observations were taken mostly in the $I$-band with additional $\approx 10\%$ \mbox{$V$-band} images to secure the color information. The MOA group uses a custom red filter, which is effectively the sum of the standard $R$ and $I$ filters.

OGLE-2013-BLG-0132 was discovered by the OGLE Early Warning System \citep{udalski2003} on 2013 March 3 at equatorial coordinates of R.A. = \ra{17}{59}{03}{51}, Decl. = \dec{-28}{25}{15}{7} (J2000.0) or Galactic coordinates $l=1.944^{\circ}$, $b=-2.275^{\circ}$. The event was independently found by the MOA group as MOA-2013-BLG-148 on 2013 March 13.

OGLE-2013-BLG-1721 was discovered through the ``new'' object channel\footnote{Events discovered through the ``new'' object channel are too faint in the baseline to be detected on the deep reference image of a given field.} by the OGLE Early Warning System on 2013 August 30. The event is located at equatorial coordinates of R.A. = \ra{17}{52}{30}{37}, Decl. = \dec{-30}{17}{33}{7} (J2000.0), i.e., $l=-0.393^{\circ}$, $b=-1.981^{\circ}$. MOA announced the same event as MOA-2013-BLG-618 on 2013 September 2. The event was also observed by the Wise group \citep{yossi2016} and it was included in their statistical analysis of the frequency of snow line-region planets.

For the final analysis, the OGLE and MOA data sets were re-reduced. Both events occurred on faint stars and during the peak magnification, they did not exceed $I\approx 17$. The OGLE light curve of OGLE-2013-BLG-0132 shows a long-term trend in the baseline, due to a high proper motion of a neighboring star. The trend was removed prior to further modeling. The MOA data show more systematics due to faintness of events, worse weather conditions, and pixel scale. We carefully detrended the data, but we decided to use 20-day subsets of light curves, which cover peaks and planetary anomalies of both events. This choice has a small effect on the best-fitting model parameters. OGLE photometric uncertainties were re-scaled following the algorithm of \citet{skowron2016}. For the MOA data set $\chi^2/{\rm dof}\sim 1$ and therefore we did not apply error corrections.

\begin{figure}[h]
\centering
\includegraphics[width=0.5\textwidth]{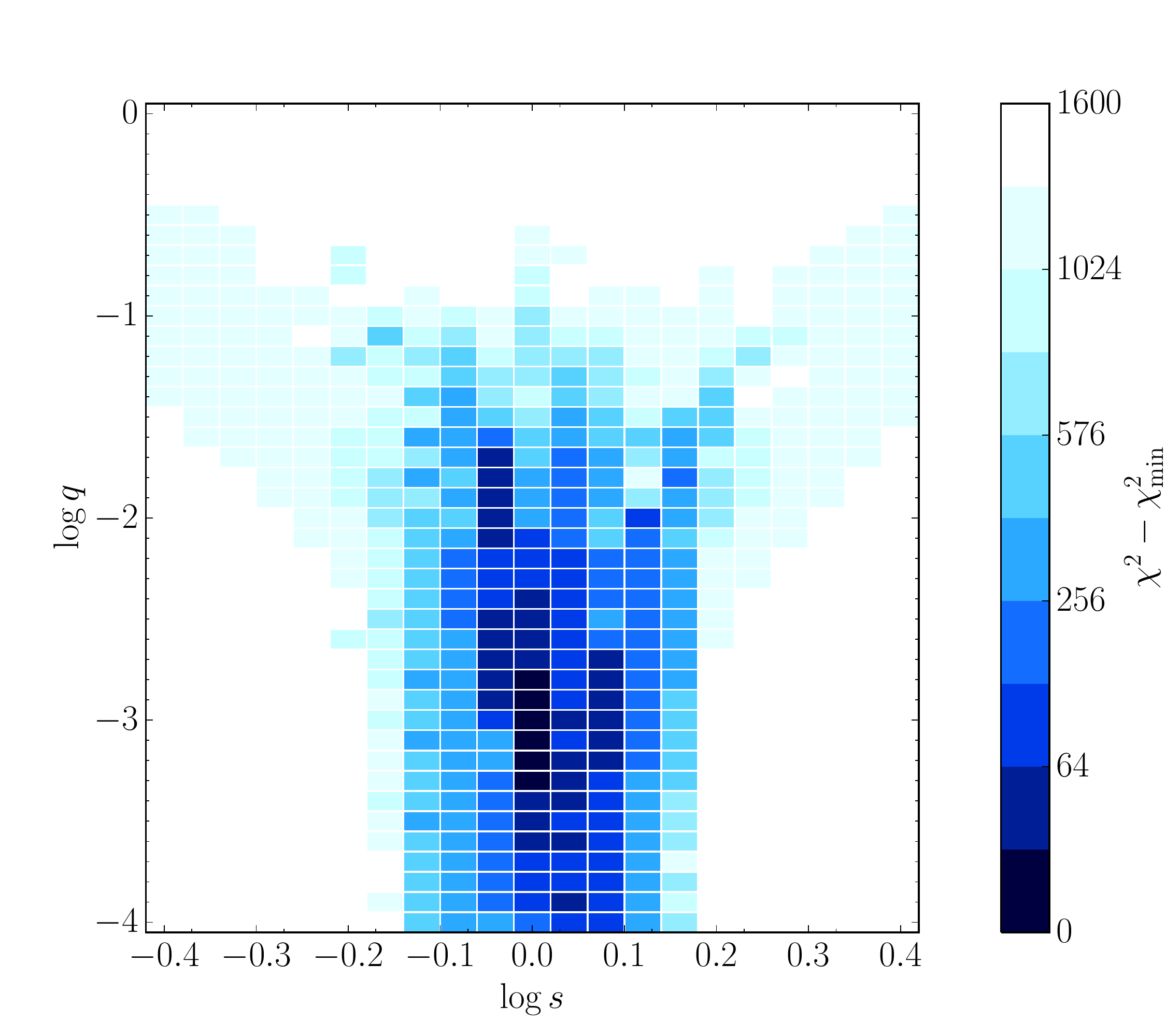}
\caption{$\Delta\chi^2$ map in the $\log{q}\!-\!\log{s}$ parameter space obtained from the preliminary grid search for OGLE-2013-BLG-0132. The strongest local minimum is located around $(\log{q},\log{s})=(-3.1,0.0)$}
\label{fig:grid1}
\end{figure}

\begin{figure}[h]
\centering
\includegraphics[width=0.5\textwidth]{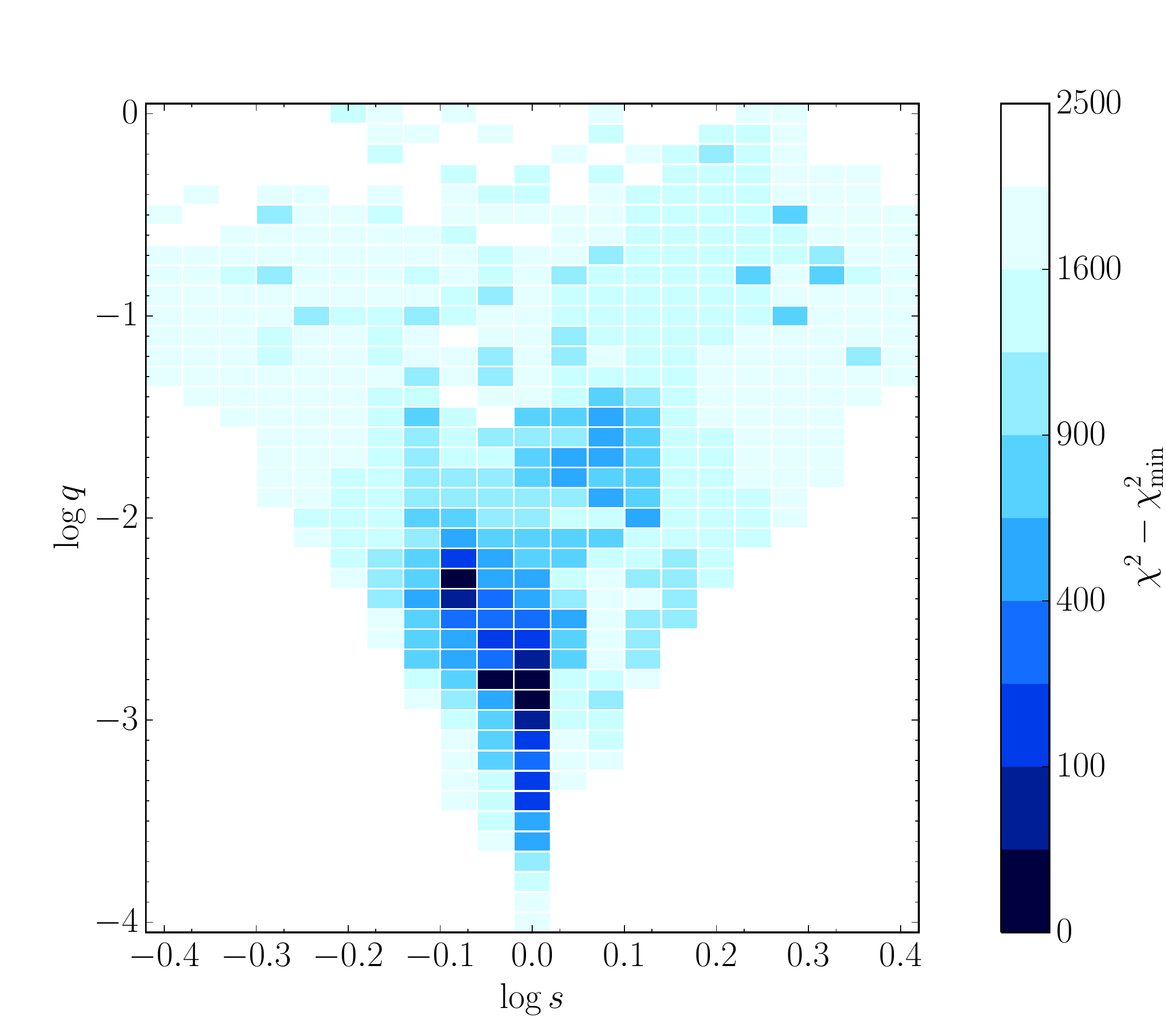}
\caption{$\Delta\chi^2$ map in the $\log{q}\!-\!\log{s}$ parameter space obtained from the preliminary grid search for OGLE-2013-BLG-1721. There are two strong local minima around $(\log{q},\log{s})=(-2.8,0.0)$ and $(-2.3,-0.08)$. Subsequent MCMC investigations in these areas reveal deeper minimum in the first region, with $\Delta\chi^2=248$ when compared to the best minimum in the second region.}
\label{fig:grid2}
\end{figure}

\begin{table}
\caption{Best-fit Model Parameters}
\begin{tabular}{lr@{$\pm$}lr@{$\pm$}l}
\hline
Parameters & \multicolumn{2}{c}{OGLE-2013-BLG-0132} & \multicolumn{2}{c}{OGLE-2013-BLG-1721}  \\
\hline
$\chi^2/{\rm dof}$ & \multicolumn{2}{c}{1104/1019} & \multicolumn{2}{c}{1682/1544} \\
$t_0$ (HJD$'$)          & 6370.156 & 0.064  & 6535.280 & 0.022 \\
$u_0$                 & 0.184 & 0.005     & 0.060 & 0.003 \\
$t_{\rm E}$ (days)    & 36.99 & 0.77      & 27.98 & 0.65 \\
$q\ (10^{-4})$        & 5.15 & 0.28       & 13.18 & 0.72 \\
$s$                   & 1.150 & 0.004     & 0.964 & 0.002 \\
$\alpha$ (rad)        & 0.821 & 0.008     & $-0.781$ & 0.017 \\
$\rho\ (10^{-3})$ & 1.0 & 0.1         & 1.2 & 0.1 \\
\hline
$I_{\rm s}$           & 19.37 & 0.03    & 21.27 & 0.04 \\
$f_{\rm s}$           & 0.663 & 0.018     & 0.552 & 0.018 \\
\hline
\end{tabular}
Note. HJD$'$=HJD-2450000. $f_{\rm s} = F_{\rm s} / (F_{\rm s} + F_{\rm b})$ is the blending parameter.
\label{tab:fit}
\end{table}

\section{Light curves modeling}
\label{sec:model}

Light curves of both events can be described by seven parameters. Four of them characterize the source trajectory relative to the lens: $t_0$ and $u_0$ are the time and projected distance (in Einstein radius units) during the closest approach to the lens center of mass, $t_{\rm E}$ is the Einstein radius crossing time, and $\alpha$ is the angle between the source-lens relative trajectory and the binary axis. The binary lens is described by two parameters: the mass ratio $q=M_2/M_1$ and projected separation $s$ (in units of the angular Einstein radius $\theta_{\rm E}$). An additional parameter, $\rho$, the normalized source angular radius, is needed to take into account the finite-source effect. 

Two additional parameters for each observatory describe the source flux $F_{\rm s}$ and blended, unmagnified flux $F_{\rm b}$ from unresolved neighbors and/or the lens itself. The modeled light curve is given by $F(t)=A(t)F_{\rm s}+F_{\rm b}$, where $F(t)$ is the flux at time $t$ and $A(t)=A(t;t_0,t_{\rm E},u_0,\rho,q,s,\alpha)$ is the magnification. For a given set of parameters $(t_0,t_{\rm E},u_0,\rho,q,s,\alpha)$, the source and blend fluxes were calculated analytically using the linear least-squares approach.

Lensing magnifications during planetary anomalies were computed using the ray-shooting method \citep{schneider1986,kayser1986,wamb1997,skowron2015}. We used the point-source approximation far from the caustics and the hexadecapole approximation \citep{gould2008,pejcha2009} at intermediate distances. 

Light curve modeling was performed in a few steps. First, we found local minima of the $\chi^2$ surface, performing a grid search with $q$, $s$, and $\alpha$ fixed at $41 \times 21\times 80$ different grid points. The ranges of grid parameters were $-4 \leq \log{q}\leq 0$, $-0.4 \leq \log{s} \leq 0.4$, and $-\pi \leq \alpha \leq \pi$. We also fixed the source angular radius $\log\rho=-3.0$. The remaining parameters were searched for using the Sequential Least SQuares Programming optimization algorithm \citep{kraft}. Results of the preliminary grid search are shown in Figures \ref{fig:grid1} and \ref{fig:grid2}. Subsequently, we explored all local minima using the Markov Chain Monte Carlo (MCMC) method. For both microlensing events, we found only one unique solution. We checked whether there are any other degenerate models using the MultiNest algorithm \citep{multi1,multi2,poleski2017}, but we found no solutions with comparable $\chi^2$.

We found that both light curves can be satisfactorily described by a seven-parameter model. The inclusion of higher-order effects (like the parallax or orbital motion) does not improve $\chi^2$ significantly ($\Delta\chi^2=2.7$ for OGLE-2013-BLG-0132 and $\Delta\chi^2=3.0$ for OGLE-2013-BLG-1721). This is not surprising, as both microlenses were short and faint, and the higher-order effects are more likely to manifest in brighter and longer events.

\begin{figure*}[h]
\centering
\includegraphics[width=0.9\textwidth]{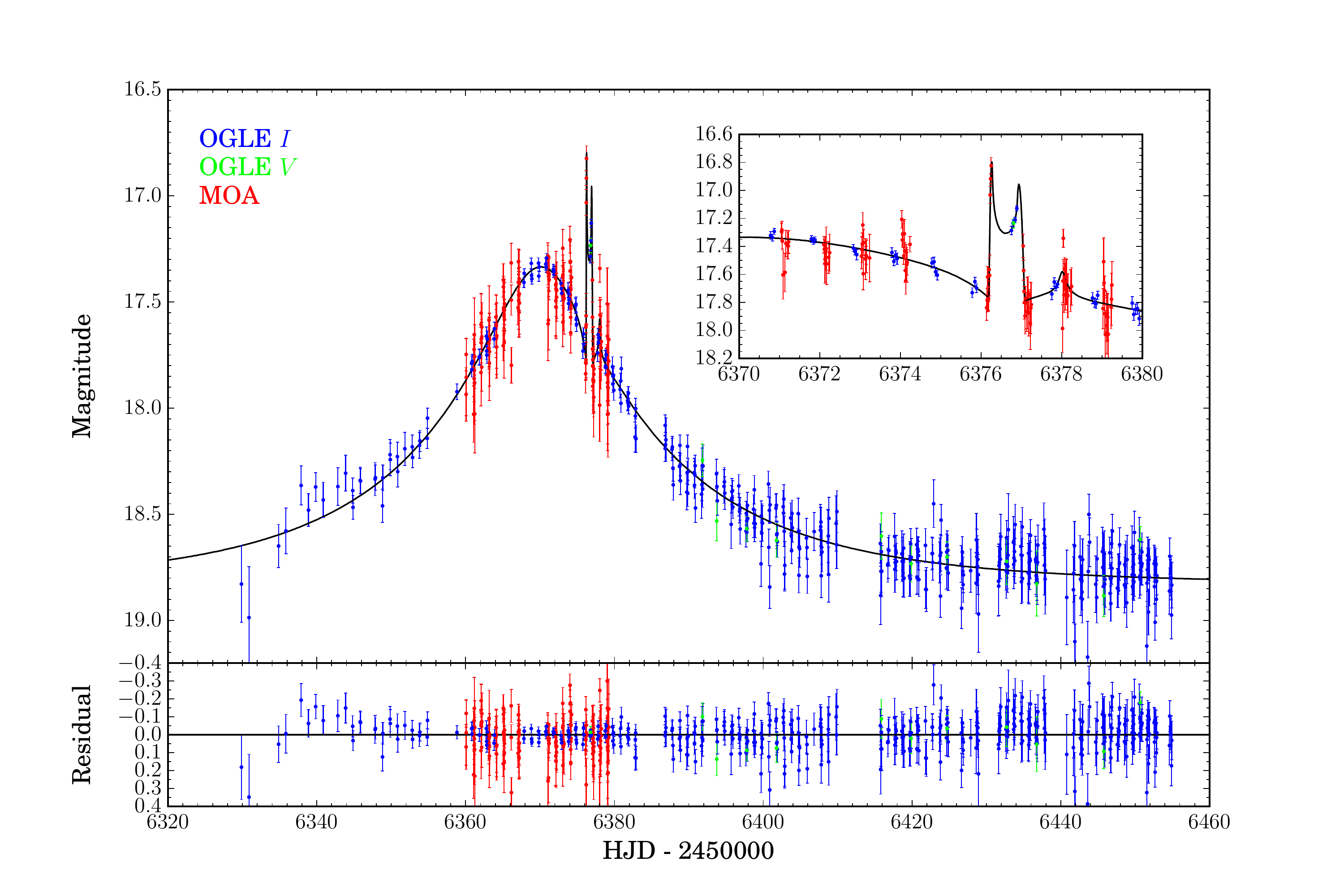}
\caption{Light curve of OGLE-2013-BLG-0132. The inset shows the enlargement of the caustic-crossing parts of the light curve. The lower panel shows the residuals from the best-fit model.}
\label{fig:132lc}
\end{figure*}

\begin{figure*}[h]
\centering
\includegraphics[width=0.9\textwidth]{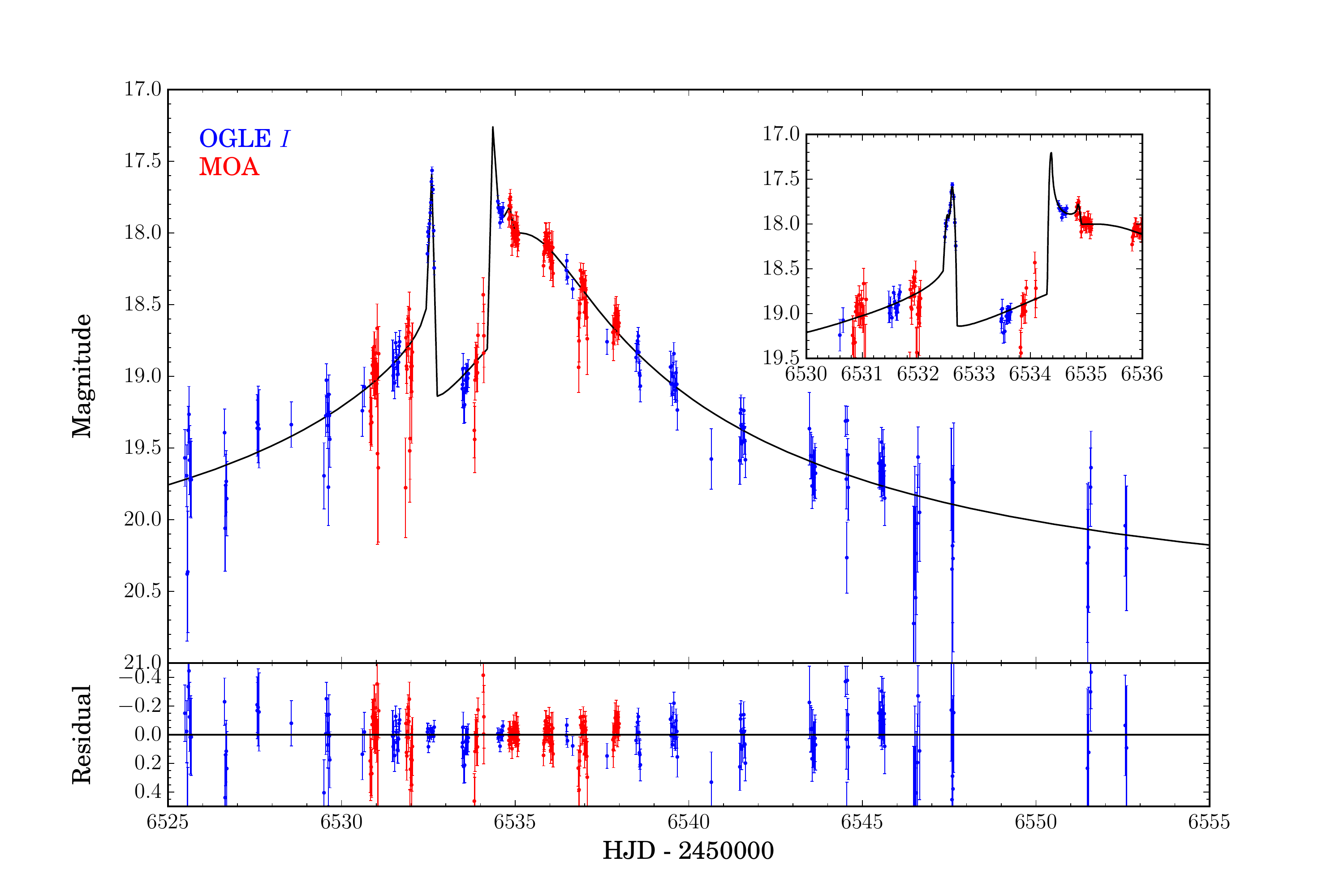}
\caption{Light curve of OGLE-2013-BLG-1721. The inset shows the enlargement of the caustic-crossing parts of the light curve. The lower panel shows the residuals from the best-fit model.}
\label{fig:1721lc}
\end{figure*}

\begin{figure}[h]
\centering
\includegraphics[width=0.4\textwidth]{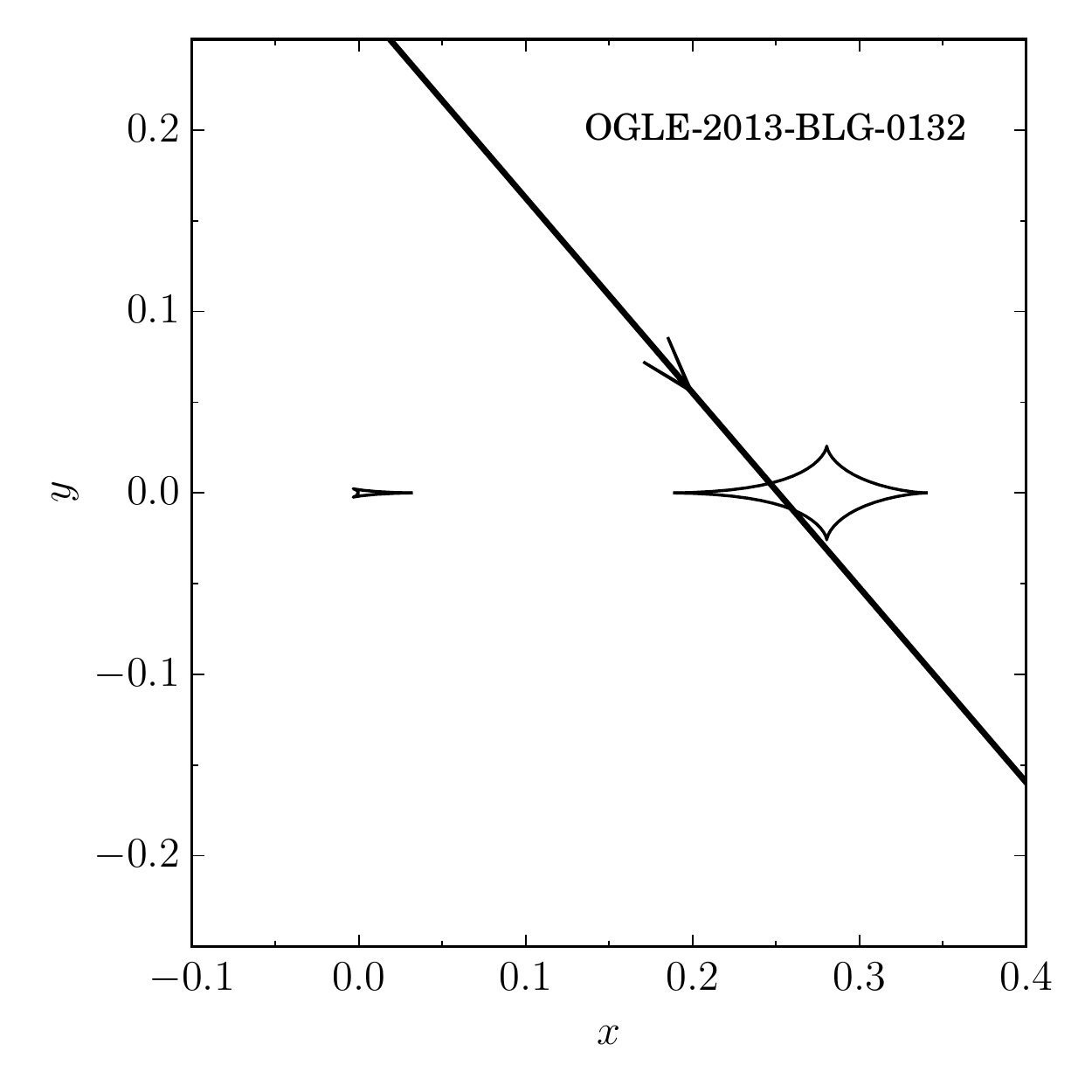}
\includegraphics[width=0.4\textwidth]{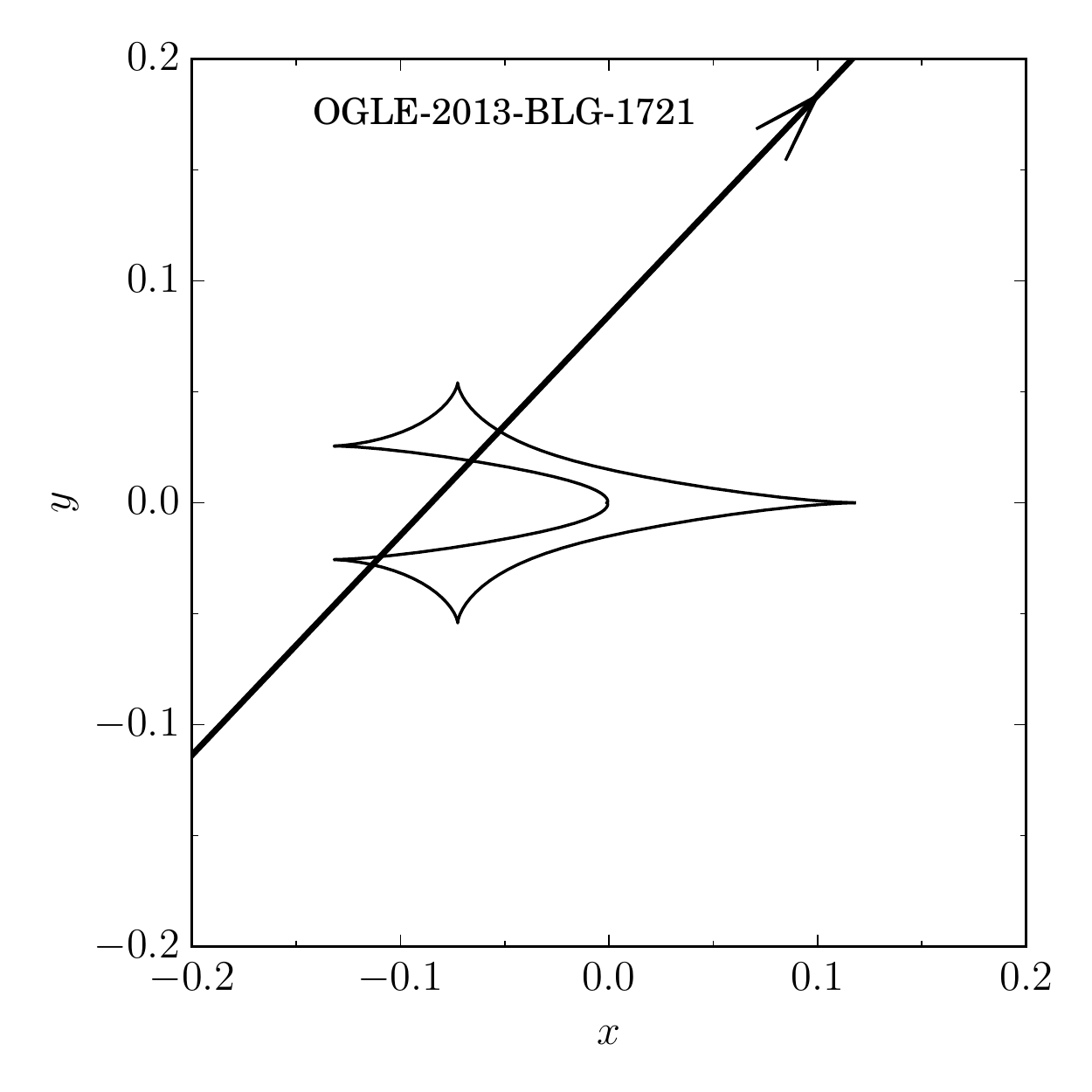}
\caption{The caustic curves corresponding to best-fitting models of OGLE-2013-BLG-0132 (upper panel) and OGLE-2013-BLG-1721 (lower panel). The source-lens relative trajectory is shown by a straight line (the direction of the source star is indicated with an arrow). Source size is smaller than a width of the line. Host stars are located near $(0,0)$ and planets are at $x=1.150$ and $x=0.964$, respectively.}
\label{fig:geom}
\end{figure}

The best-fit parameters and their $1\sigma$ uncertainties (calculated from a Markov chain) are shown in Table \ref{tab:fit}. The light curves and best-fitting models are shown in Figures \ref{fig:132lc} and \ref{fig:1721lc}. Caustic geometries are presented in Figure \ref{fig:geom}.

\section{Source stars}
\label{sec:source}

\subsection{OGLE-2013-BLG-0132}

We estimate the calibrated source brightness $I_{\rm S} = 19.37 \pm 0.03$ and color $(V-I)_{\rm S} = 1.79 \pm 0.04$ from the microlensing model by comparing $V$- and $I$-band light curves with predicted magnifications. The de-reddened brightness and color are calculated under the assumption that the source suffers the same amount of extinction as the red clump stars. We measured the red clump centroid on the calibrated color--magnitude diagram for stars in the $2' \times 2'$ field centered on the microlensing event: $I_{\rm RC} = 15.62$ and $(V-I)_{\rm RC} = 2.07$ (Figure \ref{fig:cmds}). The intrinsic color of red clump stars is $(V-I)_{\rm RC,0} = 1.06$ \citep{bensby2011} and their mean de-reddened brightness in this direction is $I_{\rm RC,0}=14.36$ \citep{nataf2013}. This yields de-reddened color and brightness of the source star: $(V-I)_{\rm S,0} = (V-I)_{\rm RC,0} - (V-I)_{\rm RC} + (V-I)_{\rm S} = 0.78 \pm 0.04$ and $I_{\rm S,0} = I_{\rm RC,0} - I_{\rm RC} + I_{\rm S} = 18.11 \pm 0.20$. 

Using the color--color relation from \citet{bessel1988}, we find $(V-K)_{\rm S,0} = 1.71 \pm 0.11$. Then, from the color--surface brightness relation for dwarf stars from \citet{kervella2004}, we estimate the angular radius of the source star $\theta_* = 0.81 \pm 0.10$ $\mu$as. This gives the angular Einstein radius $\theta_{\rm E} = \theta_* / \rho = 0.81 \pm 0.12$ mas and the relative lens-source proper motion $\mu_{\rm rel}= \theta_{\rm E} / t_{\rm E} = 8.0 \pm 1.3$ mas/yr.

\subsection{OGLE-2013-BLG-1721}

\begin{figure*}
\includegraphics[width=0.5\textwidth]{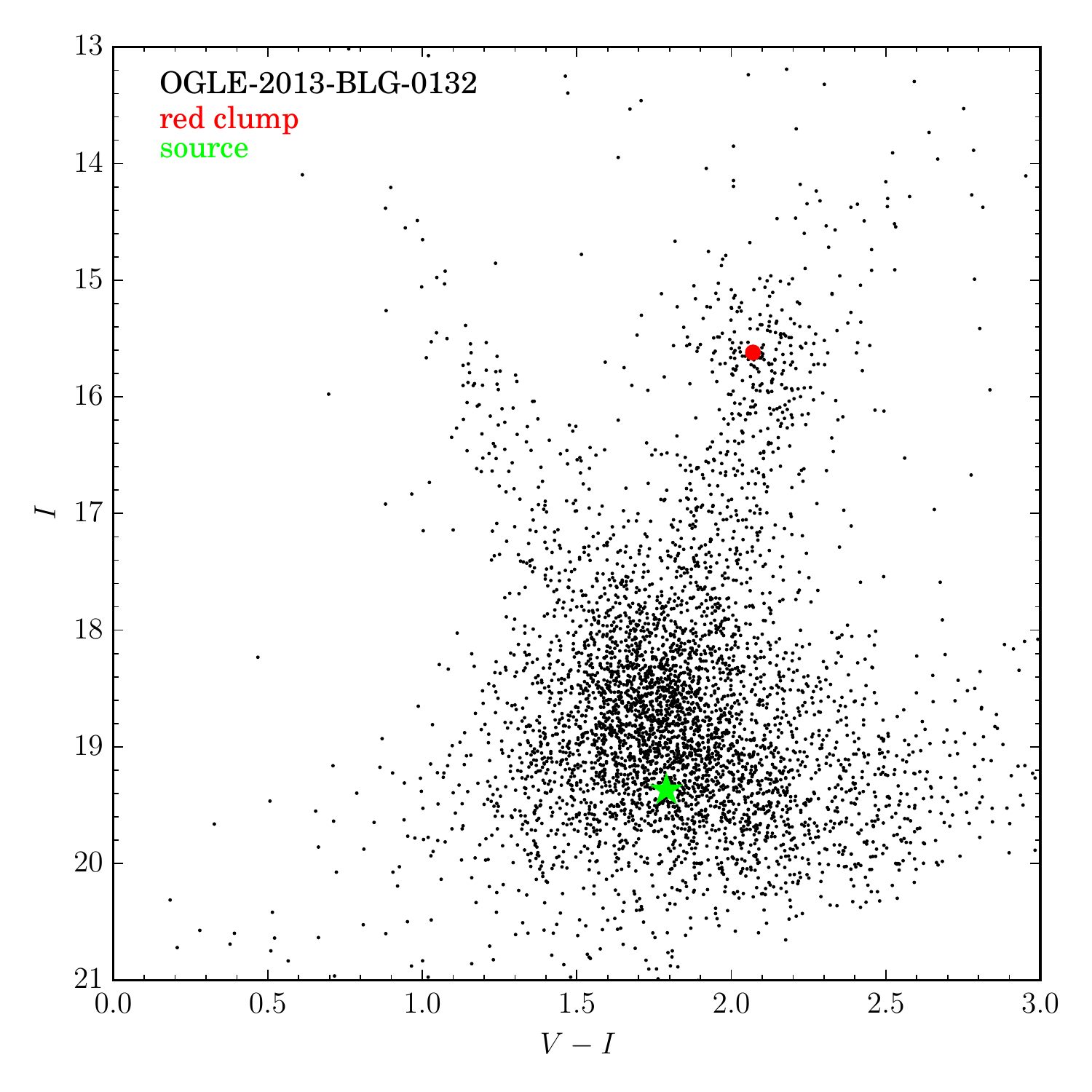}
\includegraphics[width=0.5\textwidth]{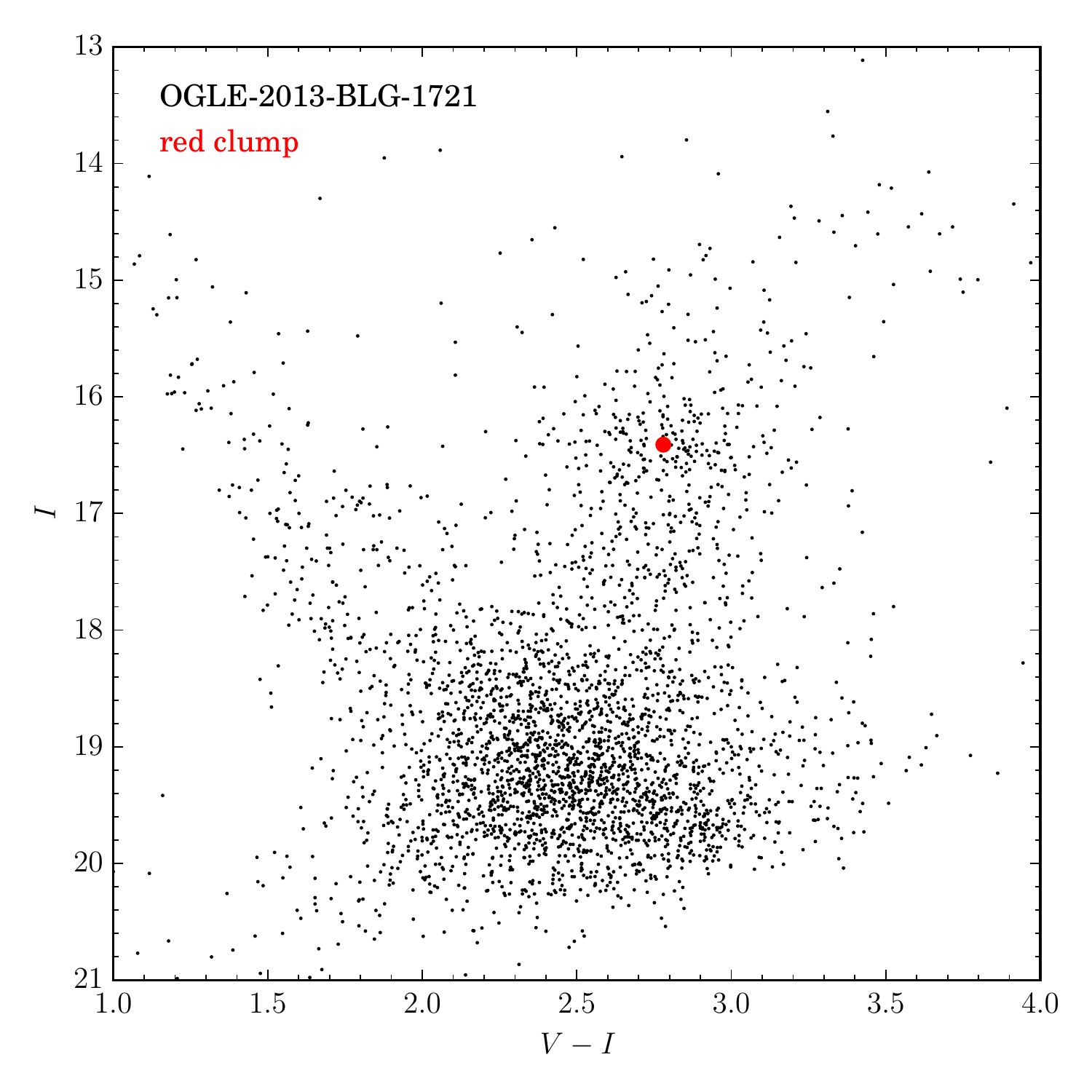}
\caption{OGLE color--magnitude diagrams of stars in the $2'\times 2'$ field centered on the microlensing event.}
\label{fig:cmds}
\end{figure*}

We estimate the calibrated source brightness of $I_{\rm S} = 21.27 \pm 0.04$ from the microlensing model. Despite the fact that the field was observed in the \mbox{$V$-band} several times, the event was not detected in OGLE images, meaning it was fainter than $V_{\rm peak} \gtrsim 21$. Therefore, we are unable to assess source properties in a standard way \citep{yoo2004}. Red clump stars are still visible in the $2'\times 2'$ field centered on the microlensing event (which corresponds to $\sim 5 \times 5$~pc at the Galactic center distance; Figure \ref{fig:cmds}). Assuming that the source is located in the Galactic bulge, we can still characterize it. We measure the red clump centroid on the calibrated color-magnitude diagram: $I_{\rm RC} = 16.41$ and $(V-I)_{\rm RC} = 2.78$. \citet{nataf2013} provide the mean de-reddened red clump brightness $I_{\rm RC,0}=14.38$ in this direction, so the reddening $E(V-I) = (V-I)_{\rm RC} -(V-I)_{\rm RC,0} = 2.78-1.06 = 1.72$ and $A_I = I_{\rm RC}-I_{\rm RC,0}=16.41-14.38 = 2.03$ and $A_V = A_I + E(V-I) =3.75$. The source in $I$-band is 4.86 mag fainter than the red clump, below the main-sequence turn-off \citep{zoccali2003}. Hence, we conclude that the source most likely lies on the main-sequence and has absolute brightness $M_I = -0.12 + 4.86 = 4.74$, which corresponds to a $0.9\ {\rm M}_{\odot}$ G9 dwarf \citep{mamajek2013}. Its $V$-band absolute brightness is $M_V = 5.55$, so $V_{\rm S} = I_{\rm S} + (V-I)_{\rm S,0} + E(V-I)=23.80$, and the source could not have been detected in OGLE $V$-band observations.

Using color--color relations from \citet{bessel1988}, we find $(V-K)_{\rm S,0} = 1.81 \pm 0.10$, so the source angular radius is $\theta_* = 0.51 \pm 0.10$ $\mu$as \citep{kervella2004}. This gives the angular Einstein radius $\theta_{\rm E} = \theta_* / \rho = 0.42 \pm 0.09$ mas and the relative lens-source proper motion $\mu_{\rm rel} = \theta_{\rm E} / t_{\rm E} = 5.5 \pm 1.2$ mas/yr.

\section{Bayesian analysis}
\label{sec:bayes}
For both events we were unable to measure the microlens parallax $\pi_{\rm E}$, which is crucial for calculating the lens mass. As it was mentioned in Section \ref{sec:model}, both events were short and faint, while the parallax is more likely to manifest in longer events. Additionally, because the signal is subtle, it requires good photometric precision. $3\sigma$ upper limits of the microlens parallax are $\pi_{\rm E} \lesssim 1.4$ for OGLE-2013-BLG-0132 and $\pi_{\rm E} \lesssim 2.6$ for OGLE-2013-BLG-1721, which provide very weak constraints on the lens mass and distance. We employ the Bayesian analysis to estimate the lens' parameters. We use the Galactic model of \citet{han2003} (see also \citealt{batista2011}). In short, they use the ``G2'' bulge model of \citet{dwek1995} and the disk model of \citet{zheng2001}. The lens-source relative proper motion is calculated based on velocity distributions of \citet{han1995}.

We incorporate the \citet{kroupa2001} initial mass function, which is approximated by a broken power law with $\alpha=0.3$ in a brown dwarf regime ($0.01<M/M_{\odot}<0.08$), $\alpha=1.3$ for $0.08<M/M_{\odot}<0.5$, and $\alpha=2.3$ for $0.5<M/M_{\odot}<150$. We assume that all stars with initial masses $1 < M/M_{\odot} < 8$ evolved into white dwarfs, and we adopt the empirical initial-final mass relation for white dwarfs $M_{\rm final} = 0.339 \pm 0.015 + (0.129\pm 0.004) M_{\rm init}$ \citep{williams2009}. Masses of neutron stars (with initial masses in the range $8 < M/M_{\odot} < 20$) concentrate around $1.33\ M_{\odot}$ \citep{kiziltan}, while for black holes we assume a Gaussian distribution at $7.8 \pm 1.2\ M_{\odot}$ \citep{ozel2010}. We emphasize that we explicitly assume the probability of hosting a planet is independent of the host mass, its evolutionary stage, and location within the Milky Way.

\begin{figure}
\centering
\includegraphics[width=0.5\textwidth]{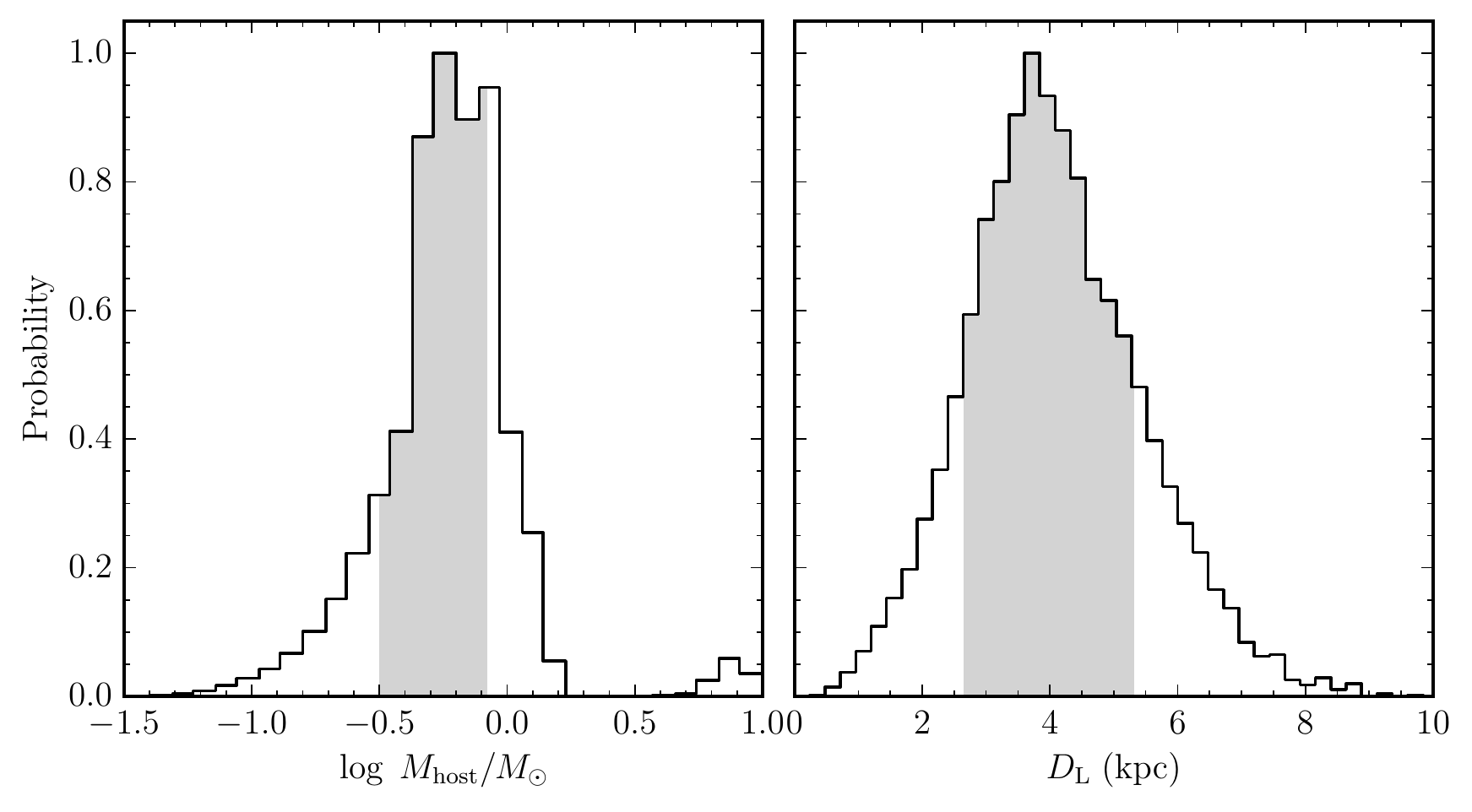}
\caption{Posterior distributions for the lens mass (left) and distance (right) calculated from the Bayesian analysis for OGLE-2013-BLG-0132. Shaded regions correspond to a $1\sigma$ confidence interval.}
\label{fig:bayes1}
\end{figure}

\begin{figure}
\centering
\includegraphics[width=0.5\textwidth]{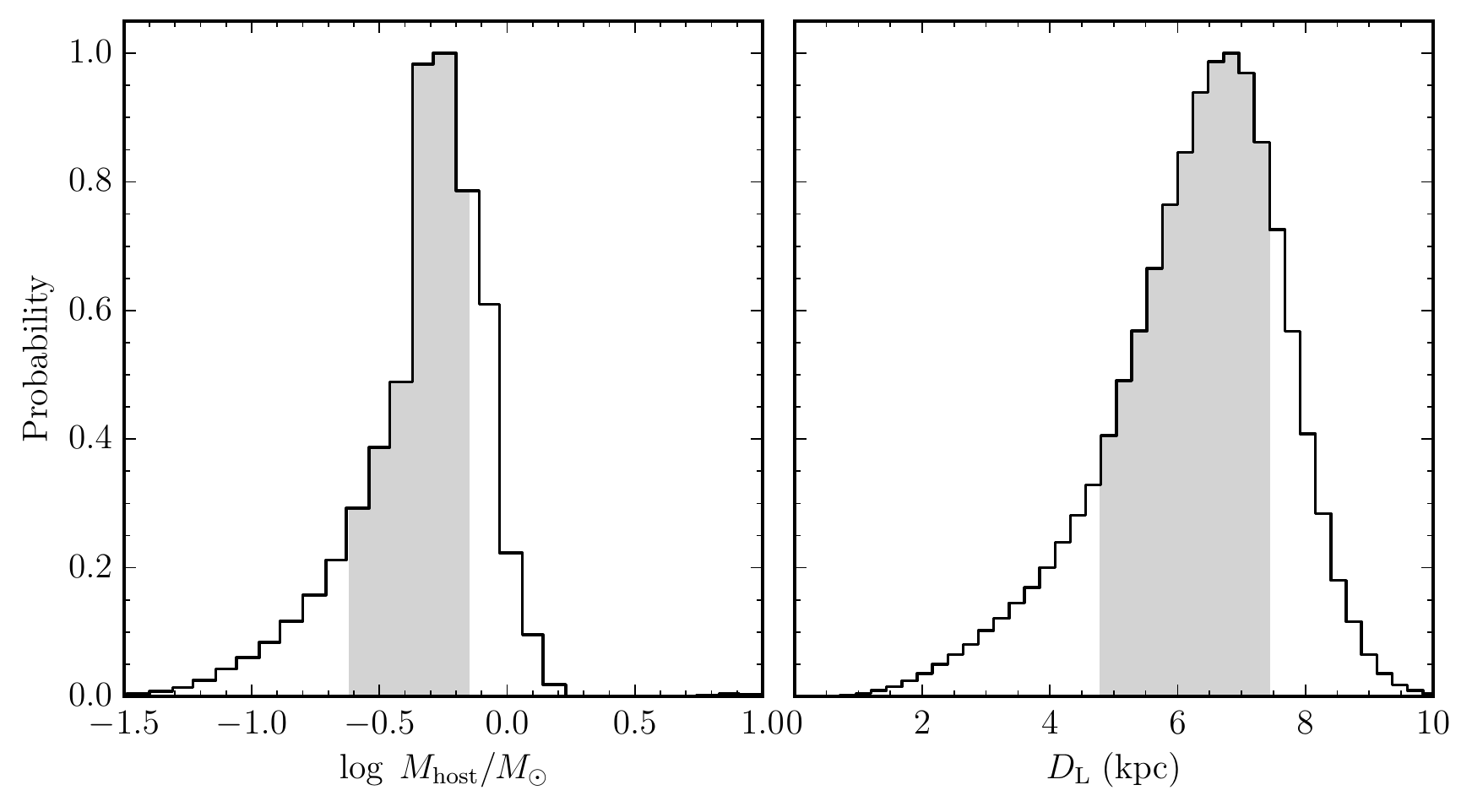}
\caption{Posterior distributions for the lens mass (left) and distance (right) calculated from the Bayesian analysis for OGLE-2013-BLG-1721. Shaded regions correspond to a $1\sigma$ confidence interval.}
\label{fig:bayes2}
\end{figure}

We use the measured values of $t_{\rm E}$ and $\theta_{\rm E}$ to constrain the lens distance and mass. Posterior distributions for these quantities are shown in Figures \ref{fig:bayes1} and \ref{fig:bayes2}. Host stars are likely early M-type dwarfs ($0.54^{+0.30}_{-0.23}\ M_{\odot}$ for OGLE-2013-BLG-0132 and $0.46^{+0.26}_{-0.23}\ M_{\odot}$ for OGLE-2013-BLG-1721), located at a distance of $3.9^{+1.5}_{-1.3}$ kpc and $6.3^{+1.1}_{-1.6}$ kpc, respectively (Table \ref{tab:phys}). At such distances, hosts would have $I$-band magnitudes of approx. 22.0 and 24.3, which are consistent with constraints from the blend flux ($I_{\rm lens} > 20.11$ for OGLE-2013-BLG-0132 and $I_{\rm lens} >21.50$ for OGLE-2013-BLG-1721). The mass of the planet is estimated to $0.29^{+0.16}_{-0.13}\ M_{\rm Jup}$ for OGLE-2013-BLG-0132 and $0.64^{+0.35}_{-0.31}\ M_{\rm Jup}$ for OGLE-2013-BLG-1721. The projected separations are $3.6^{+1.4}_{-1.2}$ and $2.6^{+0.5}_{-0.7}$ au, respectively. For the event OGLE-2013-BLG-1721, $\theta_{\rm E}$ was measured indirectly, because we lacked $V$-band observations. However, the Bayesian analysis based solely on $t_{\rm E}$ yields virtually identical results: $M_{\rm host} = 0.44^{+0.28}_{-0.25}\ M_{\odot}$ and $D_{\rm L} = 6.1^{+1.5}_{-2.2}$ kpc.

\begin{table}
\caption{Physical Parameters ($1\sigma$ confidence intervals)}
\centering
\def\arraystretch{1.5}
\begin{tabular}{lrrrr}
\hline
Parameters & \multicolumn{1}{c}{OGLE-2013-BLG-0132} & \multicolumn{1}{c}{OGLE-2013-BLG-1721}  \\
\hline
$\theta_{\rm E}$ (mas)       & $0.81 \pm 0.12$        & $0.42 \pm 0.09$         \\
$\mu_{\rm rel}$ (mas/yr)     & $8.0 \pm 1.3$          & $5.5 \pm 1.2$           \\
$M_{\rm host}$ ($M_{\odot}$) & $0.54^{+0.30}_{-0.23}$ & $0.46^{+0.26}_{-0.23}$  \\
$D_{\rm L}$ (kpc)            & $3.9^{+1.5}_{-1.3}$    & $6.3^{+1.1}_{-1.6} $    \\
$M_{\rm planet}$ ($M_{\rm Jup}$) & $0.29^{+0.16}_{-0.13}$ & $0.64^{+0.35}_{-0.31}$  \\
$a_{\perp}$ (au)             & $3.6^{+1.4}_{-1.2}$    & $2.6^{+0.5}_{-0.7}$     \\
\hline
\end{tabular}
\label{tab:phys}
\end{table}

\section{Discussion and conclusions}
\label{sec:conclusions}
We presented the discovery and characterization of two sub-Jupiter-mass planets orbiting M-dwarfs in two microlensing events OGLE-2013-BLG-0132 and OGLE-2013-BLG-1721. Both events showed clear deviations from the simple point-source point-lens model, caused by the presence of a second body with well-measured planet-to-host mass ratios of $(5.15\pm0.28)\times 10^{-4}$ and $(13.18\pm0.72)\times10^{-4}$, respectively.\footnote{We note that \citet{yossi2016} found the approximate mass ratio of $q\approx 11\times 10^{-4}$ for OGLE-2013-BLG-1721, based on a coarse grid search. They included the event in their statistical analysis of the frequency of snow line-region planets.} Events were short and faint, which prevented us from measuring a reliable parallax signal. We therefore used the Bayesian analysis to estimate the lens properties. From posterior distributions of the lens mass and distance, we estimate masses of planetary companions of $0.29^{+0.16}_{-0.13}\ M_{\rm Jup}$ (OGLE-2013-BLG-0132Lb) and $0.64^{+0.35}_{-0.31}\ M_{\rm Jup}$ (OGLE-2013-BLG-1721Lb) and their projected separations of $3.6^{+1.4}_{-1.2}$ au and $2.6^{+0.5}_{-0.7}$ au, respectively. Both planets likely belong to a group of sub-Jupiter-mass planets ($0.2 < M/M_{\rm Jup} < 1$) orbiting M-dwarfs beyond the snow line distance. It was recently suggested that such planets are very common \citep{fukui2015,hirao2016}, in agreement with predictions of the core-accretion theory of planet formation. A number of exoplanets discovered using the microlensing technique is growing very fast, and therefore, those presented here and future discoveries will allow the planet population around M-dwarfs to be constrained.

Direct observations of the lens can pinpoint the mass and distance to the system (e.g., \citealt{bennett2006,bennett2015,batista2015}). OGLE-2013-BLG-0132 is a promising candidate for the future high-resolution imaging follow-up. The relative lens-source proper motion is high ($8.0 \pm 1.3$ mas/yr), and the source and blend have similar $I$-band brightness. Follow-up observations of OGLE-2013-BLG-1721 will be more challenging. The relative proper motion is smaller ($5.5 \pm 1.2$ mas/yr) and the lens is $\sim 3$ mag fainter than the source (in the $I$-band). In the $H$-band, this difference is smaller ($\sim 2$ mag), but still significant. The pixel size of the Near-Infrared Camera on JWST is 32 mas, so the lens and source will separate by one pixel by 2017 (OGLE-2013-BLG-0132) and 2019 (OGLE-2013-BLG-1721).

\section*{Acknowledgements}

The OGLE team thanks Professors M. Kubiak and G. Pietrzy\'nski, former members of the team, for their contribution to the collection of the OGLE photometric data over the past years. The OGLE project has received funding from the National Science Center, Poland, grant MAESTRO 2014/14/A/ST9/00121 to AU. The MOA project is supported by JSPS Kakenhi grants JP24253004, JP26247023, JP16H06287, JP23340064, and JP15H00781 and by the Royal Society of New Zealand Marsden Grant MAU1104. Work by C.H. was supported by the Creative Research Initiative Program (2009-0081561) of the National Research Foundation of Korea.

\bibliographystyle{aasjournal}
\bibliography{pap}

\begin{thebibliography}{}
\expandafter\ifx\csname natexlab\endcsname\relax\def\natexlab#1{#1}\fi

\bibitem[{{Batista} {et~al.}(2015){Batista}, {Beaulieu}, {Bennett}, {Gould},
  {Marquette}, {Fukui}, \& {Bhattacharya}}]{batista2015}
{Batista}, V., {Beaulieu}, J.-P., {Bennett}, D.~P., {et~al.} 2015, \apj, 808,
  170

\bibitem[{{Batista} {et~al.}(2011){Batista}, {Gould}, {Dieters}, {Dong},
  {Bond}, {Beaulieu}, {Maoz}, {Monard}, {Christie}, {McCormick}, {Albrow},
  {Horne}, {Tsapras}, {Burgdorf}, {Calchi Novati}, {Skottfelt}, {Caldwell},
  {Koz{\l}owski}, {Kubas}, {Gaudi}, {Han}, {Bennett}, {An}, {MOA
  Collaboration}, {Abe}, {Botzler}, {Douchin}, {Freeman}, {Fukui}, {Furusawa},
  {Hearnshaw}, {Hosaka}, {Itow}, {Kamiya}, {Kilmartin}, {Korpela}, {Lin},
  {Ling}, {Makita}, {Masuda}, {Matsubara}, {Miyake}, {Muraki}, {Nagaya},
  {Nishimoto}, {Ohnishi}, {Okumura}, {Perrott}, {Rattenbury}, {Saito},
  {Sullivan}, {Sumi}, {Sweatman}, {Tristram}, {von Seggern}, {Yock}, {PLANET
  Collaboration}, {Brillant}, {Calitz}, {Cassan}, {Cole}, {Cook}, {Coutures},
  {Dominis Prester}, {Donatowicz}, {Greenhill}, {Hoffman}, {Jablonski}, {Kane},
  {Kains}, {Marquette}, {Martin}, {Martioli}, {Meintjes}, {Menzies},
  {Pedretti}, {Pollard}, {Sahu}, {Vinter}, {Wambsganss}, {Watson}, {Williams},
  {Zub}, {FUN Collaboration}, {Allen}, {Bolt}, {Bos}, {DePoy}, {Drummond},
  {Eastman}, {Gal-Yam}, {Gorbikov}, {Higgins}, {Janczak}, {Kaspi}, {Lee},
  {Mallia}, {Maury}, {Monard}, {Moorhouse}, {Morgan}, {Natusch}, {Ofek},
  {Park}, {Pogge}, {Polishook}, {Santallo}, {Shporer}, {Spector}, {Thornley},
  {Yee}, {MiNDSTEp Consortium}, {Bozza}, {Browne}, {Dominik}, {Dreizler},
  {Finet}, {Glitrup}, {Grundahl}, {Harps{\o}e}, {Hessman}, {Hinse},
  {Hundertmark}, {J{\o}rgensen}, {Liebig}, {Maier}, {Mancini}, {Mathiasen},
  {Rahvar}, {Ricci}, {Scarpetta}, {Southworth}, {Surdej}, {Zimmer}, {RoboNet
  Collaboration}, {Allan}, {Bramich}, {Snodgrass}, {Steele}, \&
  {Street}}]{batista2011}
{Batista}, V., {Gould}, A., {Dieters}, S., {et~al.} 2011, \aap, 529, A102

\bibitem[{{Bennett} {et~al.}(2006){Bennett}, {Anderson}, {Bond}, {Udalski}, \&
  {Gould}}]{bennett2006}
{Bennett}, D.~P., {Anderson}, J., {Bond}, I.~A., {Udalski}, A., \& {Gould}, A.
  2006, \apjl, 647, L171

\bibitem[{{Bennett} {et~al.}(2010){Bennett}, {Rhie}, {Nikolaev}, {Gaudi},
  {Udalski}, {Gould}, {Christie}, {Maoz}, {Dong}, {McCormick}, {Szyma{\'n}ski},
  {Tristram}, {Macintosh}, {Cook}, {Kubiak}, {Pietrzy{\'n}ski},
  {Soszy{\'n}ski}, {Szewczyk}, {Ulaczyk}, {Wyrzykowski}, {OGLE Collaboration},
  {DePoy}, {Han}, {Kaspi}, {Lee}, {Mallia}, {Natusch}, {Park}, {Pogge},
  {Polishook}, {{$\mu$}FUN Collaboration}, {Abe}, {Bond}, {Botzler}, {Fukui},
  {Hearnshaw}, {Itow}, {Kamiya}, {Korpela}, {Kilmartin}, {Lin}, {Ling},
  {Masuda}, {Matsubara}, {Motomura}, {Muraki}, {Nakamura}, {Okumura},
  {Ohnishi}, {Perrott}, {Rattenbury}, {Sako}, {Saito}, {Sato}, {Skuljan},
  {Sullivan}, {Sumi}, {Sweatman}, {Yock}, {MOA Collaboration}, {Albrow},
  {Allan}, {Beaulieu}, {Bramich}, {Burgdorf}, {Coutures}, {Dominik}, {Dieters},
  {Fouqu{\'e}}, {Greenhill}, {Horne}, {Snodgrass}, {Steele}, {Tsapras},
  {PLANET}, {RoboNet Collaborations}, {Chaboyer}, {Crocker}, \&
  {Frank}}]{bennett2010}
{Bennett}, D.~P., {Rhie}, S.~H., {Nikolaev}, S., {et~al.} 2010, \apj, 713, 837

\bibitem[{{Bennett} {et~al.}(2015){Bennett}, {Bhattacharya}, {Anderson},
  {Bond}, {Anderson}, {Barry}, {Batista}, {Beaulieu}, {DePoy}, {Dong}, {Gaudi},
  {Gilbert}, {Gould}, {Pfeifle}, {Pogge}, {Suzuki}, {Terry}, \&
  {Udalski}}]{bennett2015}
{Bennett}, D.~P., {Bhattacharya}, A., {Anderson}, J., {et~al.} 2015, \apj, 808,
  169

\bibitem[{{Bensby} {et~al.}(2011){Bensby}, {Ad{\'e}n}, {Mel{\'e}ndez}, {Gould},
  {Feltzing}, {Asplund}, {Johnson}, {Lucatello}, {Yee}, {Ram{\'{\i}}rez},
  {Cohen}, {Thompson}, {Bond}, {Gal-Yam}, {Han}, {Sumi}, {Suzuki}, {Wada},
  {Miyake}, {Furusawa}, {Ohmori}, {Saito}, {Tristram}, \&
  {Bennett}}]{bensby2011}
{Bensby}, T., {Ad{\'e}n}, D., {Mel{\'e}ndez}, J., {et~al.} 2011, \aap, 533,
  A134

\bibitem[{{Bessell} \& {Brett}(1988)}]{bessel1988}
{Bessell}, M.~S., \& {Brett}, J.~M. 1988, \pasp, 100, 1134

\bibitem[{{Bond} {et~al.}(2001){Bond}, {Abe}, {Dodd}, {Hearnshaw}, {Honda},
  {Jugaku}, {Kilmartin}, {Marles}, {Masuda}, {Matsubara}, {Muraki}, {Nakamura},
  {Nankivell}, {Noda}, {Noguchi}, {Ohnishi}, {Rattenbury}, {Reid}, {Saito},
  {Sato}, {Sekiguchi}, {Skuljan}, {Sullivan}, {Sumi}, {Takeuti}, {Watase},
  {Wilkinson}, {Yamada}, {Yanagisawa}, \& {Yock}}]{bond2001}
{Bond}, I.~A., {Abe}, F., {Dodd}, R.~J., {et~al.} 2001, \mnras, 327, 868

\bibitem[{{Dwek} {et~al.}(1995){Dwek}, {Arendt}, {Hauser}, {Kelsall}, {Lisse},
  {Moseley}, {Silverberg}, {Sodroski}, \& {Weiland}}]{dwek1995}
{Dwek}, E., {Arendt}, R.~G., {Hauser}, M.~G., {et~al.} 1995, \apj, 445, 716

\bibitem[{{Feroz} \& {Hobson}(2008)}]{multi1}
{Feroz}, F., \& {Hobson}, M.~P. 2008, \mnras, 384, 449

\bibitem[{{Feroz} {et~al.}(2009){Feroz}, {Hobson}, \& {Bridges}}]{multi2}
{Feroz}, F., {Hobson}, M.~P., \& {Bridges}, M. 2009, \mnras, 398, 1601

\bibitem[{{Fukui} {et~al.}(2015){Fukui}, {Gould}, {Sumi}, {Bennett}, {Bond},
  {Han}, {Suzuki}, {Beaulieu}, {Batista}, {Udalski}, {Street}, {Tsapras},
  {Hundertmark}, {Abe}, {Bhattacharya}, {Freeman}, {Itow}, {Ling}, {Koshimoto},
  {Masuda}, {Matsubara}, {Muraki}, {Ohnishi}, {Philpott}, {Rattenbury},
  {Saito}, {Sullivan}, {Tristram}, {Yonehara}, {MOA Collaboration}, {Choi},
  {Christie}, {DePoy}, {Dong}, {Drummond}, {Gaudi}, {Hwang}, {Kavka}, {Lee},
  {McCormick}, {Natusch}, {Ngan}, {Park}, {Pogge}, {Shin}, {Tan}, {Yee},
  {{$\mu$}FUN Collaboration}, {Szyma{\'n}ski}, {Pietrzy{\'n}ski},
  {Soszy{\'n}ski}, {Poleski}, {Koz{\l}owski}, {Pietrukowicz}, {Ulaczyk},
  {Wyrzykowski}, {OGLE Collaboration}, {Bramich}, {Browne}, {Dominik}, {Horne},
  {Ipatov}, {Kains}, {Snodgrass}, {Steele}, \& {RoboNet
  Collaboration}}]{fukui2015}
{Fukui}, A., {Gould}, A., {Sumi}, T., {et~al.} 2015, \apj, 809, 74

\bibitem[{{Gaudi} {et~al.}(2008){Gaudi}, {Bennett}, {Udalski}, {Gould},
  {Christie}, {Maoz}, {Dong}, {McCormick}, {Szyma{\'n}ski}, {Tristram},
  {Nikolaev}, {Paczy{\'n}ski}, {Kubiak}, {Pietrzy{\'n}ski}, {Soszy{\'n}ski},
  {Szewczyk}, {Ulaczyk}, {Wyrzykowski}, {OGLE Collaboration}, {DePoy}, {Han},
  {Kaspi}, {Lee}, {Mallia}, {Natusch}, {Pogge}, {Park}, {{$\mu$}-Fun
  Collabortion}, {Abe}, {Bond}, {Botzler}, {Fukui}, {Hearnshaw}, {Itow},
  {Kamiya}, {Korpela}, {Kilmartin}, {Lin}, {Masuda}, {Matsubara}, {Motomura},
  {Muraki}, {Nakamura}, {Okumura}, {Ohnishi}, {Rattenbury}, {Sako}, {Saito},
  {Sato}, {Skuljan}, {Sullivan}, {Sumi}, {Sweatman}, {Yock}, {MOA
  Collaboration}, {Albrow}, {Allan}, {Beaulieu}, {Burgdorf}, {Cook},
  {Coutures}, {Dominik}, {Dieters}, {Fouqu{\'e}}, {Greenhill}, {Horne},
  {Steele}, {Tsapras}, {Planet Collaboration}, {RoboNet Collaborations},
  {Chaboyer}, {Crocker}, {Frank}, \& {Macintosh}}]{gaudi2008}
{Gaudi}, B.~S., {Bennett}, D.~P., {Udalski}, A., {et~al.} 2008, Science, 319,
  927

\bibitem[{{Gould}(2008)}]{gould2008}
{Gould}, A. 2008, \apj, 681, 1593

\bibitem[{{Han} \& {Gould}(1995)}]{han1995}
{Han}, C., \& {Gould}, A. 1995, \apj, 447, 53

\bibitem[{{Han} \& {Gould}(2003)}]{han2003}
---. 2003, \apj, 592, 172

\bibitem[{{Hirao} {et~al.}(2016){Hirao}, {Udalski}, {Sumi}, {Bennett}, {Bond},
  {Rattenbury}, {Suzuki}, {Koshimoto}, {Abe}, {Asakura}, {Bhattacharya},
  {Freeman}, {Fukui}, {Itow}, {Li}, {Ling}, {Masuda}, {Matsubara}, {Matsuo},
  {Muraki}, {Nagakane}, {Ohnishi}, {Oyokawa}, {Saito}, {Sharan}, {Shibai},
  {Sullivan}, {Tristram}, {Yonehara}, {MOA Collaboration}, {Poleski},
  {Skowron}, {Mr{\'o}z}, {Szyma{\'n}ski}, {Koz{\l}owski}, {Pietrukowicz},
  {Soszy{\'n}ski}, {Wyrzykowski}, {Ulaczyk}, \& {OGLE
  Collaboration}}]{hirao2016}
{Hirao}, Y., {Udalski}, A., {Sumi}, T., {et~al.} 2016, \apj, 824, 139

\bibitem[{{Ida} \& {Lin}(2005)}]{ida2005}
{Ida}, S., \& {Lin}, D.~N.~C. 2005, \apj, 626, 1045

\bibitem[{{Kains} {et~al.}(2013){Kains}, {Street}, {Choi}, {Han}, {Udalski},
  {Almeida}, {Jablonski}, {Tristram}, {J{\o}rgensen}, {Szyma{\'n}ski},
  {Kubiak}, {Pietrzy{\'n}ski}, {Soszy{\'n}ski}, {Poleski}, {Koz{\l}owski},
  {Pietrukowicz}, {Ulaczyk}, {Wyrzykowski}, {Skowron}, {Alsubai}, {Bozza},
  {Browne}, {Burgdorf}, {Calchi Novati}, {Dodds}, {Dominik}, {Dreizler},
  {Fang}, {Grundahl}, {Gu}, {Hardis}, {Harps{\o}e}, {Hessman}, {Hinse},
  {Hornstrup}, {Hundertmark}, {Jessen-Hansen}, {Kerins}, {Liebig}, {Lund},
  {Lundkvist}, {Mancini}, {Mathiasen}, {Penny}, {Rahvar}, {Ricci}, {Sahu},
  {Scarpetta}, {Skottfelt}, {Snodgrass}, {Southworth}, {Surdej},
  {Tregloan-Reed}, {Wambsganss}, {Wertz}, {Bajek}, {Bramich}, {Horne},
  {Ipatov}, {Steele}, {Tsapras}, {Abe}, {Bennett}, {Bond}, {Botzler}, {Chote},
  {Freeman}, {Fukui}, {Furusawa}, {Itow}, {Ling}, {Masuda}, {Matsubara},
  {Miyake}, {Muraki}, {Ohnishi}, {Rattenbury}, {Saito}, {Sullivan}, {Sumi},
  {Suzuki}, {Suzuki}, {Sweatman}, {Takino}, {Wada}, {Yock}, {Allen}, {Batista},
  {Chung}, {Christie}, {DePoy}, {Drummond}, {Gaudi}, {Gould}, {Henderson},
  {Jung}, {Koo}, {Lee}, {McCormick}, {McGregor}, {Mu{\~n}oz}, {Natusch},
  {Ngan}, {Park}, {Pogge}, {Shin}, {Yee}, {Albrow}, {Bachelet}, {Beaulieu},
  {Brillant}, {Caldwell}, {Cassan}, {Cole}, {Corrales}, {Coutures}, {Dieters},
  {Dominis Prester}, {Donatowicz}, {Fouqu{\'e}}, {Greenhill}, {Kane}, {Kubas},
  {Marquette}, {Martin}, {Meintjes}, {Menzies}, {Pollard}, {Williams},
  {Wouters}, \& {Zub}}]{kains2013}
{Kains}, N., {Street}, R.~A., {Choi}, J.-Y., {et~al.} 2013, \aap, 552, A70

\bibitem[{{Kayser} {et~al.}(1986){Kayser}, {Refsdal}, \&
  {Stabell}}]{kayser1986}
{Kayser}, R., {Refsdal}, S., \& {Stabell}, R. 1986, \aap, 166, 36

\bibitem[{{Kennedy} \& {Kenyon}(2008)}]{kennedy2008}
{Kennedy}, G.~M., \& {Kenyon}, S.~J. 2008, \apj, 673, 502

\bibitem[{{Kervella} {et~al.}(2004){Kervella}, {Th{\'e}venin}, {Di Folco}, \&
  {S{\'e}gransan}}]{kervella2004}
{Kervella}, P., {Th{\'e}venin}, F., {Di Folco}, E., \& {S{\'e}gransan}, D.
  2004, \aap, 426, 297

\bibitem[{{Kiziltan} {et~al.}(2013){Kiziltan}, {Kottas}, {De Yoreo}, \&
  {Thorsett}}]{kiziltan}
{Kiziltan}, B., {Kottas}, A., {De Yoreo}, M., \& {Thorsett}, S.~E. 2013, \apj,
  778, 66

\bibitem[{{Kraft}(1988)}]{kraft}
{Kraft}, D. 1988, {A software package for sequential quadratic programming},
  Tech. Rep. DFVLR-FB 88-28, DLR German Aerospace Center -- Institute for
  Flight Mechanics, Koln, Germany

\bibitem[{{Kroupa}(2001)}]{kroupa2001}
{Kroupa}, P. 2001, \mnras, 322, 231

\bibitem[{{Laughlin} {et~al.}(2004){Laughlin}, {Bodenheimer}, \&
  {Adams}}]{laughlin2004}
{Laughlin}, G., {Bodenheimer}, P., \& {Adams}, F.~C. 2004, \apjl, 612, L73

\bibitem[{{Mizuno}(1980)}]{mizuno1980}
{Mizuno}, H. 1980, Progress of Theoretical Physics, 64, 544

\bibitem[{{Nataf} {et~al.}(2013){Nataf}, {Gould}, {Fouqu{\'e}}, {Gonzalez},
  {Johnson}, {Skowron}, {Udalski}, {Szyma{\'n}ski}, {Kubiak},
  {Pietrzy{\'n}ski}, {Soszy{\'n}ski}, {Ulaczyk}, {Wyrzykowski}, \&
  {Poleski}}]{nataf2013}
{Nataf}, D.~M., {Gould}, A., {Fouqu{\'e}}, P., {et~al.} 2013, \apj, 769, 88

\bibitem[{{{\"O}zel} {et~al.}(2010){{\"O}zel}, {Psaltis}, {Narayan}, \&
  {McClintock}}]{ozel2010}
{{\"O}zel}, F., {Psaltis}, D., {Narayan}, R., \& {McClintock}, J.~E. 2010,
  \apj, 725, 1918

\bibitem[{{Pecaut} \& {Mamajek}(2013)}]{mamajek2013}
{Pecaut}, M.~J., \& {Mamajek}, E.~E. 2013, \apjs, 208, 9

\bibitem[{{Pejcha} \& {Heyrovsk{\'y}}(2009)}]{pejcha2009}
{Pejcha}, O., \& {Heyrovsk{\'y}}, D. 2009, \apj, 690, 1772

\bibitem[{{Poleski} {et~al.}(2017){Poleski}, {Udalski}, {Bond}, {Beaulieu},
  {Clanton}, {Gaudi}, {Szyma{\'n}ski}, {Soszy{\'n}ski}, {Pietrukowicz},
  {Koz{\l}owski}, {Skowron}, {Wyrzykowski}, {Ulaczyk}, {Bennett}, {Sumi},
  {Suzuki}, {Rattenbury}, {Koshimoto}, {Abe}, {Asakura}, {Barry},
  {Bhattacharya}, {Donachie}, {Evans}, {Fukui}, {Hirao}, {Itow}, {Li}, {Ling},
  {Masuda}, {Matsubara}, {Muraki}, {Nagakane}, {Ohnishi}, {Ranc}, {Saito},
  {Sharan}, {Sullivan}, {Tristram}, {Yamada}, {Yamada}, {Yonehara}, {Batista},
  \& {Marquette}}]{poleski2017}
{Poleski}, R., {Udalski}, A., {Bond}, I.~A., {et~al.} 2017, \aap, 604, A103

\bibitem[{{Pollack} {et~al.}(1996){Pollack}, {Hubickyj}, {Bodenheimer},
  {Lissauer}, {Podolak}, \& {Greenzweig}}]{pollack1996}
{Pollack}, J.~B., {Hubickyj}, O., {Bodenheimer}, P., {et~al.} 1996, \icarus,
  124, 62

\bibitem[{{Rattenbury} {et~al.}(2015){Rattenbury}, {Bennett}, {Sumi},
  {Koshimoto}, {Bond}, {Udalski}, {Abe}, {Bhattacharya}, {Freeman}, {Fukui},
  {Itow}, {Li}, {Ling}, {Masuda}, {Matsubara}, {Muraki}, {Ohnishi}, {Saito},
  {Sharan}, {Sullivan}, {Suzuki}, {Tristram}, {Koz{\l}owski}, {Mr{\'o}z},
  {Pietrukowicz}, {Pietrzy{\'n}ski}, {Poleski}, {Skowron}, {Skowron},
  {Soszy{\'n}ski}, {Szyma{\'n}ski}, {Ulaczyk}, \&
  {Wyrzykowski}}]{rattenbury2015}
{Rattenbury}, N.~J., {Bennett}, D.~P., {Sumi}, T., {et~al.} 2015, \mnras, 454,
  946

\bibitem[{{Sako} {et~al.}(2008){Sako}, {Sekiguchi}, {Sasaki}, {Okajima}, {Abe},
  {Bond}, {Hearnshaw}, {Itow}, {Kamiya}, {Kilmartin}, {Masuda}, {Matsubara},
  {Muraki}, {Rattenbury}, {Sullivan}, {Sumi}, {Tristram}, {Yanagisawa}, \&
  {Yock}}]{sako2008}
{Sako}, T., {Sekiguchi}, T., {Sasaki}, M., {et~al.} 2008, Experimental
  Astronomy, 22, 51

\bibitem[{{Schneider} \& {Weiss}(1986)}]{schneider1986}
{Schneider}, P., \& {Weiss}, A. 1986, \aap, 164, 237

\bibitem[{{Shvartzvald} {et~al.}(2016){Shvartzvald}, {Maoz}, {Udalski}, {Sumi},
  {Friedmann}, {Kaspi}, {Poleski}, {Szyma{\'n}ski}, {Skowron}, {Koz{\l}owski},
  {Wyrzykowski}, {Mr{\'o}z}, {Pietrukowicz}, {Pietrzy{\'n}ski},
  {Soszy{\'n}ski}, {Ulaczyk}, {Abe}, {Barry}, {Bennett}, {Bhattacharya},
  {Bond}, {Freeman}, {Inayama}, {Itow}, {Koshimoto}, {Ling}, {Masuda}, {Fukui},
  {Matsubara}, {Muraki}, {Ohnishi}, {Rattenbury}, {Saito}, {Sullivan},
  {Suzuki}, {Tristram}, {Wakiyama}, \& {Yonehara}}]{yossi2016}
{Shvartzvald}, Y., {Maoz}, D., {Udalski}, A., {et~al.} 2016, \mnras, 457, 4089

\bibitem[{{Skowron} {et~al.}(2015){Skowron}, {Shin}, {Udalski}, {Han}, {Sumi},
  {Shvartzvald}, {Gould}, {Dominis Prester}, {Street}, {J{\o}rgensen},
  {Bennett}, {Bozza}, {Szyma{\'n}ski}, {Kubiak}, {Pietrzy{\'n}ski},
  {Soszy{\'n}ski}, {Poleski}, {Koz{\l}owski}, {Pietrukowicz}, {Ulaczyk},
  {Wyrzykowski}, {OGLE Collaboration}, {Abe}, {Bhattacharya}, {Bond},
  {Botzler}, {Freeman}, {Fukui}, {Fukunaga}, {Itow}, {Ling}, {Koshimoto},
  {Masuda}, {Matsubara}, {Muraki}, {Namba}, {Ohnishi}, {Philpott},
  {Rattenbury}, {Saito}, {Sullivan}, {Suzuki}, {Tristram}, {Yock}, {MOA
  Collaboration}, {Maoz}, {Kaspi}, {Friedmann}, {Wise Group}, {Almeida},
  {Batista}, {Christie}, {Choi}, {DePoy}, {Gaudi}, {Henderson}, {Hwang},
  {Jablonski}, {Jung}, {Lee}, {McCormick}, {Natusch}, {Ngan}, {Park}, {Pogge},
  {Yee}, {{$\mu$}FUN Collaboration}, {Albrow}, {Bachelet}, {Beaulieu},
  {Brillant}, {Caldwell}, {Cassan}, {Cole}, {Corrales}, {Coutures}, {Dieters},
  {Donatowicz}, {Fouqu{\'e}}, {Greenhill}, {Kains}, {Kane}, {Kubas},
  {Marquette}, {Martin}, {Menzies}, {Pollard}, {Ranc}, {Sahu}, {Wambsganss},
  {Williams}, {Wouters}, {PLANET Collaboration}, {Tsapras}, {Bramich}, {Horne},
  {Hundertmark}, {Snodgrass}, {Steele}, {RoboNet Collaboration}, {Alsubai},
  {Browne}, {Burgdorf}, {Calchi Novati}, {Dodds}, {Dominik}, {Dreizler},
  {Fang}, {Gu}, {Hardis}, {Harps{\o}e}, {Hessman}, {Hinse}, {Hornstrup},
  {Jessen-Hansen}, {Kerins}, {Liebig}, {Lund}, {Lundkvist}, {Mancini},
  {Mathiasen}, {Penny}, {Rahvar}, {Ricci}, {Scarpetta}, {Skottfelt},
  {Southworth}, {Surdej}, {Tregloan-Reed}, {Wertz}, \& {MiNDSTEp
  Consortium}}]{skowron2015}
{Skowron}, J., {Shin}, I.-G., {Udalski}, A., {et~al.} 2015, \apj, 804, 33

\bibitem[{{Skowron} {et~al.}(2016){Skowron}, {Udalski}, {Koz{\l}owski},
  {Szyma{\'n}ski}, {Mr{\'o}z}, {Wyrzykowski}, {Poleski}, {Pietrukowicz},
  {Ulaczyk}, {Pawlak}, \& {Soszy{\'n}ski}}]{skowron2016}
{Skowron}, J., {Udalski}, A., {Koz{\l}owski}, S., {et~al.} 2016, \actaa, 66, 1

\bibitem[{{Sumi} {et~al.}(2013){Sumi}, {Bennett}, {Bond}, {Abe}, {Botzler},
  {Fukui}, {Furusawa}, {Itow}, {Ling}, {Masuda}, {Matsubara}, {Muraki},
  {Ohnishi}, {Rattenbury}, {Saito}, {Sullivan}, {Suzuki}, {Sweatman},
  {Tristram}, {Wada}, {Yock}, \& {MOA Collaboratoin}}]{sumi2013}
{Sumi}, T., {Bennett}, D.~P., {Bond}, I.~A., {et~al.} 2013, \apj, 778, 150

\bibitem[{{Udalski}(2003)}]{udalski2003}
{Udalski}, A. 2003, \actaa, 53, 291

\bibitem[{{Udalski} {et~al.}(2015){Udalski}, {Szyma{\'n}ski}, \&
  {Szyma{\'n}ski}}]{udalski2015}
{Udalski}, A., {Szyma{\'n}ski}, M.~K., \& {Szyma{\'n}ski}, G. 2015, \actaa, 65,
  1

\bibitem[{{Wambsganss}(1997)}]{wamb1997}
{Wambsganss}, J. 1997, \mnras, 284, 172

\bibitem[{{Williams} {et~al.}(2009){Williams}, {Bolte}, \&
  {Koester}}]{williams2009}
{Williams}, K.~A., {Bolte}, M., \& {Koester}, D. 2009, \apj, 693, 355

\bibitem[{{Yoo} {et~al.}(2004){Yoo}, {DePoy}, {Gal-Yam}, {Gaudi}, {Gould},
  {Han}, {Lipkin}, {Maoz}, {Ofek}, {Park}, {Pogge}, {Mu-Fun Collaboration},
  {Udalski}, {Soszy{\'n}ski}, {Wyrzykowski}, {Kubiak}, {Szyma{\'n}ski},
  {Pietrzy{\'n}ski}, {Szewczyk}, {{\.Z}ebru{\'n}}, \& {OGLE
  Collaboration}}]{yoo2004}
{Yoo}, J., {DePoy}, D.~L., {Gal-Yam}, A., {et~al.} 2004, \apj, 603, 139

\bibitem[{{Zheng} {et~al.}(2001){Zheng}, {Flynn}, {Gould}, {Bahcall}, \&
  {Salim}}]{zheng2001}
{Zheng}, Z., {Flynn}, C., {Gould}, A., {Bahcall}, J.~N., \& {Salim}, S. 2001,
  \apj, 555, 393

\bibitem[{{Zoccali} {et~al.}(2003){Zoccali}, {Renzini}, {Ortolani}, {Greggio},
  {Saviane}, {Cassisi}, {Rejkuba}, {Barbuy}, {Rich}, \& {Bica}}]{zoccali2003}
{Zoccali}, M., {Renzini}, A., {Ortolani}, S., {et~al.} 2003, \aap, 399, 931

\end{thebibliography}

\end{document}